\documentclass[12pt]{article}
\newcommand{\text}{\rm}

\begin{document}

\title{\textbf{The action of }$\mathbf{N=4}$\textbf{\ Super Yang-Mills from a
chiral primary operator}}
\author{V.E.R.Lemes$^{\text{(a)}}$, M.S.Sarandy$^{\text{(b)}}$, S.P.Sorella$^{\text{%
(a)}}$ , \and A.Tanzini$^{\text{(c)}}$ and O.S.Ventura$^{\text{(a,d)}}$ \\
\\
{\small{$^{\text{(a)}}$\textit{\ UERJ, Universidade do Estado do Rio de Janeiro}}}
\\
{\small{\textit{\ Rua S\~{a}o Francisco Xavier 524, 20550-013 Maracan\~{a}, Rio de Janeiro, Brazil.}}}\\
\\
{\small{\textit{\ }$^{\text{(b)}}\,$\textit{CBPF, Centro Brasileiro de Pesquisas
F\'{\i}sicas}}}\\
{\small{\textit{\ Rua Xavier Sigaud 150, 22290-180 Urca, Rio de Janeiro, Brazil.}}}\\
\\
{\small{\textit{\ }$^{\text{(c)}}$\textit{\ Dipartimento di Fisica, Universit\'{a}
di Roma \ ''Tor Vergata ''}}}\\
{\small{\textit{Via della Ricerca Scientifica 1, 00173 Roma, Italy }}}\\
\\
{\small{$^{\text{(d)}}$\textit{Funda\c {c}\~{a}o Os\'{o}rio, Rua Paula Ramos 52,}}}\\
{\small{\textit{20264-210 Rio Comprido, Rio de Janeiro, Brazil.}}}\\
}
\maketitle

\begin{abstract}
\noindent
Using the Vafa-Witten twisted version of $N=4$ Super Yang-Mills
a subset of the supercharges actually relevant for the nonrenormalization
properties of the theory is identified. In particular, a relationship
between the gauge-fixed action and the chiral primary operator 
$tr\phi ^{2\text{ }}$ is worked out. This result can be understood as an off-shell
extension of the reduction formula introduced by Intriligator in \cite{KI}.

\setcounter{page}{0}\thispagestyle{empty}
\end{abstract}

\vfill\newpage\ \makeatother

\renewcommand{\theequation}{\thesection.\arabic{equation}}

\section{Introduction}

It is a well established fact that the $N=4$ Super Yang-Mills (SYM) theory
in four dimensions displays a set of remarkable properties, both at
perturbative and nonperturbative level. Its $\beta $-function is found to
vanish to all orders of perturbation theory \cite{Sohnius,White1}: a feature
understood as a consequence of the absence of anomalies for the
super-conformal invariance \cite{White1}. Also, the theory is believed to
exhibit an exact electromagnetic duality \cite{Olive}, not destroyed by the
quantum corrections.

More recently, the conjectured $AdS/CFT$ correspondence \cite{Maldacena}
between type $IIB$ superstring on $AdS_5\times S^5$ and $N=4$ SYM theory has
renewed the interest on the finiteness properties of this theory in its
superconformal phase. In fact, many of the tests of the $AdS/CFT$ conjecture
have relied on nonrenormalization properties, which are crucial in order to
ensure a meaningful comparison between the strong coupling regime,
accessible by type $IIB$ supergravity computations, and the weak coupling
one, where the field theory techniques are reliable. In particular, it has
been pointed out \cite{freedman} that a whole class of certain $n$-point
correlation functions of chiral primary operators should obey
nonrenormalization theorems. The analysis of these correlators provides thus
a highly nontrivial check for the $AdS/CFT$ correspondence, being, at
present, object of intensive research \cite{Bianchi,skiba,Sokatchev}.

In this work we study the $N=4$ SYM by using the Vafa-Witten twisted version 
\cite{Vafa}. As it is well known, when the model is formulated on $R^4$, the
twist simply amounts to a linear change of variables, so that the twisted
theory is completely equivalent to the conventional one. Nonetheless, the
use of the twisted variables considerably simplifies the analysis of the
finiteness properties, allowing to identify a subset of supercharges which
is actually relevant to control the ultraviolet behavior.

Moreover, the combined use of the Wess-Zumino gauge and of the
Batalin-Vilkovisky procedure will allow us to obtain a generalized BRST
operator which encodes all relevant generators of $N=4$, yielding an
algebraic off-shell characterization of the fully quantized action.

In particular, we shall be able to establish an interesting relationship
between the gauge-fixed $N=4$ SYM action and the chiral primary operator $%
tr\,\phi ^2$, where in the $N=2$ formalism $\phi $ corresponds to the scalar
field of the vector multiplet. This gauge invariant polynomial can be
regarded as a kind of perturbative prepotential for the $N=4$ SYM action,
which is in fact obtained from $tr\,\phi ^2$ by repeated application of the
twisted generators of the $N=4$ superalgebra$.$ 

This result can be understood as an off-shell extension of the reduction formula discussed for
the first time in \cite{KI}\footnote{%
See eq.(5.5) of hep-th/9811047.} and exploited in \cite{skiba,Sokatchev} in order
to analyse the nonrenormalization properties of the above mentioned
correlators\footnote{Observe that an off-shell version of the reduction formula has been obtained by \cite{Sokatchev} in the framework of the $N=2$ harmonic superspace formalism 
for $N=4$ SYM.}.

We remark that an analogous relationship has been also proven to hold in the
case of $N=2$ SYM theory \cite{Blasi}, where it has been possible to give a
complete algebraic proof of the celebrated nonrenormalization theorem of the 
$N=2$ beta function by making use of the vanishing of the anomalous
dimension of $tr\,\phi ^2.$ This property is maintained in the case of $N=4,$
providing an elementary proof of the ultraviolet finiteness of the theory.

The organization of the paper is as follows. In section $2$ we briefly
review the twisting procedure of Vafa-Witten. In section $3$ the quantum
extension of the theory is provided within the antifield formalism. The
section $4$ is devoted to the relationship between $tr\,\phi ^2$ and the
action of $N=4$. In section $5$ we summarize our main results, presenting
the conclusions.

\section{The twisted $N=4$ Super Yang-Mills theory}

The global symmetry group of $N=4$ SYM theory in euclidean space-time is $%
SU(2)_L\times SU(2)_R\times SU(4)$, where $SU(2)_L\times SU(2)_R$ is the
rotation group and $SU(4)$ the internal symmetry group of $N=4$. Hence the
twist operation can be performed in more than one way, depending on how the
internal symmetry group is combined with the rotation group \cite{Yamron}.
In the case of the twist proposed by Vafa and Witten \cite{Vafa} the $SU(4)$
is splitted as $SU(2)_F\times SU(2)_I$, so that the twisted global symmetry
group turns out to be $SU(2)_L^{\prime }\times SU(2)_R\times SU(2)_F,$ where 
$SU(2)_L^{\prime }=\mathrm{diag}\left( SU(2)_L\oplus SU(2)_I\right) $ and $%
SU(2)_F$ is a residual internal symmetry group. The fields of the $N=4$
multiplet are given by $(A_\mu $,$\lambda _u^\alpha $,$\overline{\lambda }_{%
\dot{\alpha}}^u,\Phi _{uv})$, where $(u,v=1,..,4)$ are indices of the
fundamental representation of $SU(4)$, and the six real scalar fields of the
model are collected into the antisymmetric and self-conjugate tensor $\Phi
_{uv}$. Under the twisted group, these fields decompose as
\begin{eqnarray}
A_\mu  &\rightarrow &A_\mu \;,  \nonumber  \label{1-t} \\
\overline{\lambda }_{\dot{\alpha}}^u\; &\rightarrow &\psi _\mu ^i\;, 
\nonumber \\
\lambda _u^\alpha  &\rightarrow &\eta ^i,\;\chi _{\mu \nu }^i\;,  \nonumber
\\
\Phi _{uv} &\rightarrow &B_{\mu \nu },\;\phi ^{ij}\;,  \label{1-t}
\end{eqnarray}
where $(i,j=1,2)\;$are indices of the residual isospin group $SU(2)_F$, $%
\phi ^{ij}$ is a symmetric tensor, and $\chi _{\mu \nu }^i$,$\;B_{\mu \nu }$
are self-dual with respect to the Lorentz indices. Since in our analysis
manifest isospin invariance is not needed, we further explicit the $SU(2)_F$
doublets as $\psi _\mu ^i=(\psi _\mu ,\chi _\mu )$, $\eta ^i=(\eta ,\xi
),\;\chi _{\mu \nu }^i=(\chi _{\mu \nu },\psi _{\mu \nu })\;$and the triplet
as $\phi ^{ij}=(\phi ,\overline{\phi },\tau )$. As we shall see later on,
the subset $(A_\mu ,\psi _\mu ,\chi _{\mu \nu },\eta ,\phi ,\overline{\phi }%
)\;$can be readily recognized as the twisted vector multiplet of $N=2$ \cite
{witten}.

Following \cite{lozano}, the action of $N=4$ SYM in terms of the twisted
variables is found to be
\begin{eqnarray}
S_{N=4} &=&\frac 1{g^2}tr\int d^4x\left( \frac {}{} D_\mu \phi D^\mu 
\overline{\phi }\,+\,i\psi _\mu D_\nu \chi ^{\mu \nu }\,+\,i\chi _\mu D_\nu
\psi ^{\mu \nu }-\,\chi _\mu D^\mu \xi \, \right. \nonumber \\
&&+\,\psi _\mu D^\mu \eta \mathrm{\,}-\,i\overline{\phi }\left\{ \psi _{\mu
\nu },\psi ^{\mu \nu }\right\} \,+\,i\phi \left\{ \chi _{\mu \nu },\chi
^{\mu \nu }\right\} \,+\,i\tau \left\{ \psi _{\mu \nu },\chi ^{\mu \nu
}\right\}  \nonumber \\
&&-\left\{ \psi _{\mu \nu },\chi ^{\mu \rho }\right\} B_\rho
^{\,\,\,\,\,\nu }-i\chi _{\mu \nu }\left[ \xi ,B^{\mu \nu }\right] -i\psi
_{\mu \nu }\left[ \eta ,B^{\mu \nu }\right] +4i\overline{\phi }\left\{ \xi
,\xi \right\}  \nonumber \\
&&-\,\,4i\,\phi \,\left\{ \eta ,\eta \right\} \,\,+\,\,4i\,\tau \,\left\{
\xi ,\eta \right\} \,+\,\psi _\mu \,\left[ \chi _\nu ,B^{\mu \nu }\right]
\,\,+\,\,i\,\phi \,\left\{ \chi _\mu \,,\chi ^\mu \right\}  \nonumber \\
&&-i\overline{\phi }\,\left\{ \psi _\mu ,\psi ^\mu \right\} -\,i\psi _\mu
\,\left[ \chi ^\mu ,\tau \right] \,-4\,\left[ \phi ,\overline{\phi }\right]
\,\left[ \phi ,\overline{\phi }\right] \,+4\left[ \phi ,\tau \right] \left[ 
\overline{\phi },\tau \right]  \nonumber \\
&&+[\phi \,,B_{\mu \nu }]\,\,[\overline{\phi }\,,B^{\mu \nu }]\,-\,H^\mu
\,\left( \,H_\mu \,-\,D_\mu \tau \,+\,i\,D^\nu B_{\mu \nu }\,\right) 
\nonumber  \label{invaction} \\
&&\left. +H^{\mu \nu }\left( -H_{\mu \nu }^{+}+\frac i4F_{\mu \nu
}^{+}-\frac 12\left[ B_{\mu \rho },B_{\,\,\,\nu }^\rho \right] -i\left[
B_{\mu \nu },\tau \right] \right) \right) \,,    \label{invaction}
\end{eqnarray}
where $g$ is the unique coupling constant and $H_{\mu \nu }$, $H_\mu $ are
auxiliary fields, with $H_{\mu \nu }$ self-dual and $F_{\mu \nu }^{+}=F_{\mu
\nu }+\frac 12\varepsilon _{\mu \nu \rho \sigma }F^{\rho \sigma }$.

Notice that in this formulation the invariance under the Cartan subgroup of $%
SU(2)_F$ is still preserved, so that we can define a conserved charge,
called here G-charge. The G-charges of all fields as well as their canonical
dimensions and statistics$\footnote{%
The statistics (nature) of the fields is determined by their G-charges. Even
values of the G-charge correspond to commuting (c) fields and odd values to
anticommuting (a) fields.}$ are displayed in the following table.
\begin{eqnarray*}
&& 
\begin{tabular}{|l|l|l|l|l|l|l|l|l|l|l|l|l|l|}
\hline
field & $A_\mu $ & $\xi $ & $\eta $ & $\psi _{\mu \nu }$ & $\chi _{\mu \nu }$
& $\psi _\mu $ & $\chi _\mu $ & $\phi $ & $\tau $ & $\overline{\phi }$ & $%
B_{\mu \nu }$ & $H_\mu $ & $H_{\mu \nu }$ \\ \hline
G-ch. & \thinspace 0 & \thinspace \thinspace 1 & -1 & \thinspace \thinspace 1
& -1 & \thinspace \thinspace 1 & -1 & 2 & 0 & -2 & \thinspace \thinspace 0 & 
\thinspace 0 & \thinspace 0 \\ \hline
dim. & \thinspace 1 & $\,\,\frac 32$ & $\,\,\frac 32$ & $\,\,\frac 32$ & $%
\,\,\frac 32$ & $\,\,\frac 32$ & $\,\,\frac 32$ & 1 & 1 & \thinspace
\thinspace 1 & \thinspace \thinspace 1 & \thinspace 2 & \thinspace 2 \\ 
\hline
nature & \thinspace c & \thinspace a & \thinspace a & \thinspace \thinspace a
& \thinspace \thinspace a & \thinspace \thinspace a & \thinspace \thinspace a
& c & c & \thinspace \thinspace c & \thinspace \thinspace c & \thinspace c & 
\thinspace c \\ \hline
\end{tabular}
\\
&\,\,\,\,&_{\,\,\,\,\,\,\,\,\,\,\,\,\,\,\,\,\,\,\,\,\,\mathrm{%
Table\,1.\,G-charge,\,canonical\,\,dimension\,\,and\,\,nature\,\,of\,\,the\,%
\,fields.\,}}
\end{eqnarray*}
We observe that, for the fields $(A_\mu ,\psi _\mu ,\chi _{\mu \nu },\eta
,\phi ,\overline{\phi }),$ the G-charge coincides with the R-charge of $N=2$ 
\cite{witten}.

The action $\left( \ref{invaction}\right) $ is invariant under gauge
transformations with infinitesimal parameter $\zeta $:
\begin{eqnarray}
\delta _\zeta ^gA_\mu  &=&-D_\mu \zeta =-\left( \partial _\mu \zeta +i\left[
A_\mu ,\zeta \right] \right) ,  \nonumber  \label{gaugetransf} \\
\,\delta _\zeta ^g\gamma  &=&i\left[ \zeta ,\gamma \right] \;\mathrm{with\,\,%
}\gamma =\left( \phi ,\overline{\phi },\psi _\mu ,\psi _{\mu \nu },\chi _\mu
,\chi _{\mu \nu },\xi ,\eta ,B_{\mu \nu },\tau ,H_\mu ,H_{\mu \nu }\right) 
\mathrm{\;.}  \nonumber  \label{gaugetransf} \\
&&  \label{gaugetransf}
\end{eqnarray}
Concerning the generators $(\delta _u^\alpha ,\overline{\delta }_{\dot{\alpha%
}}^u)$ of the $N=4$ superalgebra, it turns out that the twisting procedure
gives rise to the following twisted charges \cite{lozano}: two scalars, $%
\delta ^{+}$ and $\delta ^{-},$ two vectors, $\delta _\mu ^{+}$ and $\delta
_\mu ^{-}$, and two self-dual tensors $\delta _{\mu \nu }^{+}$ and $\delta
_{\mu \nu }^{-}.$ Of course, all twisted generators leave the action $\left( 
\ref{invaction}\right) $ invariant. In particular, the action of the scalar
generator $\delta ^{+}$  is
\begin{eqnarray}
&&
\begin{tabular}{ll}
$\delta ^{+}A_\mu =\psi _\mu \,,$ & $\delta ^{+}\tau =\xi $ \\ 
$\delta ^{+}\psi _\mu =D_\mu \phi \,,\,$ & $\delta ^{+}\chi _\mu =H_\mu \,,$
\\ 
$\delta ^{+}\phi =0\,\,$ & $\delta ^{+}\xi =i\left[ \tau ,\phi \right] $ \\ 
$\,\delta ^{+}\overline{\phi }=-\eta $ & $\delta ^{+}B_{\mu \nu }=\psi _{\mu
\nu }\,,$ \\ 
$\delta ^{+}\eta =i\left[ \phi ,\stackrel{\_}{\phi }\right] $ & $\delta
^{+}\psi _{\mu \nu }=i[B_{\mu \nu },\phi ]\,$ \\ 
$\delta ^{+}\chi _{\mu \nu }=H_{\mu \nu }$ & $\delta ^{+}H_\mu =i\left[ \chi
_\mu ,\phi \right] $ \\ 
$\delta ^{+}H_{\mu \nu }=i\left[ \chi _{\mu \nu },\phi \right] .$ & 
\end{tabular}
\nonumber \\
&& \label{ScTransf+}
\end{eqnarray}
As anticipated, in the first column of eq.$\left( \text{\ref{ScTransf+}}%
\right) $ we recognize the scalar transformations of the twisted $N=2$
subalgebra \cite{witten}, in presence of the auxiliary field $H_{\mu \nu }.$
For $\delta ^{-}$ one gets 
\begin{eqnarray}
\! &&
\begin{tabular}{ll}
$\delta ^{-}A_\mu =\chi _\mu \,,$ & $\delta ^{-}\tau =-\eta \,,$ \\ 
$\delta ^{-}\chi _\mu =-D_\mu \overline{\phi }\,,$ & $\delta ^{-}\psi _\mu
=-H_\mu +D_\mu \tau \,,$ \\ 
$\,\delta ^{-}\stackrel{\_}{\phi }=0\,,$ & $\delta ^{-}\eta =i\left[ \tau ,%
\stackrel{\_}{\phi }\right] ,$ \\ 
$\delta ^{-}\phi =-\xi \,\,,$ & $\delta ^{-}\chi _{\mu \nu }=i\left[ B_{\mu
\nu },\stackrel{\_}{\phi }\right] ,$ \\ 
$\delta ^{-}\xi =i\left[ \phi ,\overline{\phi }\right] \,,$ & $\delta
^{-}B_{\mu \nu }=-\chi _{\mu \nu }\,,$ \\ 
$\delta ^{-}\psi _{\mu \nu }=H_{\mu \nu }+i\left[ B_{\mu \nu },\tau \right] ,
$ & 
\end{tabular}
\nonumber \\
&&
\begin{tabular}{l}
$\delta ^{-}H_{\mu \nu }=-i\left[ \psi _{\mu \nu },\stackrel{\_}{\phi }%
\right] +i\left[ \chi _{\mu \nu },\tau \right] +i\left[ B_{\mu \nu },\eta
\right] \;,$ \\ 
$\delta ^{-}H_\mu =-D_\mu \eta +i\left[ \psi _\mu ,\stackrel{\_}{\phi }%
\right] +i\left[ \chi _\mu ,\tau \right] \,\;.$%
\end{tabular}
\nonumber \\
&&  \label{ScTransf-}
\end{eqnarray}
Analogously, for the vector transformations $\delta _\mu ^{+}$ and $\delta
_\mu ^{-}$ one obtains
\begin{eqnarray}
&&
\begin{tabular}{ll}
$\delta _\mu ^{+}A_\nu =-4i\chi _{\mu \nu }-4g_{\mu \nu }\eta \,,$ & $\delta
_\mu ^{+}\tau =\chi _\mu ,$ \\ 
$\delta _\mu ^{+}\phi =\psi _\mu \,,$ & $\delta _\mu ^{+}\stackrel{\_}{\phi }%
=0,$ \\ 
$\delta _\mu ^{+}\xi =D_\mu \tau -H_\mu \,,$ & $\delta _\mu ^{+}\eta =-D_\mu 
\overline{\phi },$ \\ 
$\delta _\mu ^{+}B_{\nu \rho }=-i\theta _{\mu \nu \rho \lambda }\chi
^\lambda \,,$ & $\delta _\mu ^{+}\psi _{\nu \rho }=D_\mu B_{\nu \rho
}+i\theta _{\mu \nu \rho \lambda }H^\lambda ,$ \\ 
$\delta _\mu ^{+}\chi _{\nu \rho }=i\theta _{\mu \nu \rho \lambda }D^\lambda 
\overline{\phi }\,,$ & $\delta _\mu ^{+}\chi _\nu =-4\left[ B_{\mu \nu },%
\overline{\phi }\right] +4ig_{\mu \nu }\left[ \tau ,\overline{\phi }\right] ,
$%
\end{tabular}
\nonumber \\
&&
\begin{tabular}{l}
$\delta _\mu ^{+}\psi _\nu =4iH_{\mu \nu }+F_{\mu \nu }-4ig_{\mu \nu }[%
\overline{\phi },\phi ]\,,$ \\ 
$\delta _\mu ^{+}H_{\nu \rho }=D_\mu \chi _{\nu \rho }+\theta _{\mu \nu \rho
\lambda }\left[ \psi ^\lambda ,\overline{\phi }\right] +i\theta _{\mu \nu
\rho \lambda }D^\lambda \eta \,,$ \\ 
$\delta _\mu ^{+}H_\nu =D_\mu \chi _\nu +4\left[ \eta ,B_{\mu \nu }\right]
+4\left[ \psi _{\mu \nu },\overline{\phi }\right] -4ig_{\mu \nu }\left[ \eta
,\tau \right] -4ig_{\mu \nu }\left[ \xi ,\overline{\phi }\right] \,,$%
\end{tabular}
\nonumber \\
&&  \label{VeTransf+}
\end{eqnarray}
and
\begin{eqnarray}
&&
\begin{tabular}{ll}
$\delta _\mu ^{-}A_\nu =-4i\psi _{\mu \nu }+4g_{\mu \nu }\xi \,,$ & $\delta
_\mu ^{-}\tau =\psi _\mu ,$ \\ 
$\delta _\mu ^{-}\phi =0\,,$ & $\delta _\mu ^{-}\stackrel{\_}{\phi }=-\chi
_\mu ,$ \\ 
$\delta _\mu ^{-}\xi =-D_\mu \phi \,,$ & $\delta _\mu ^{-}\eta =-H_\mu ,$ \\ 
$\delta _\mu ^{-}B_{\nu \rho }=+i\theta _{\mu \nu \rho \lambda }\psi
^\lambda \,,$ & $\delta _\mu ^{-}\psi _\nu =-4\left[ B_{\mu \nu },\phi
\right] -4ig_{\mu \nu }\left[ \tau ,\phi \right] ,$ \\ 
$\delta _\mu ^{-}\psi _{\nu \rho }=-i\theta _{\mu \nu \rho \lambda
}D^\lambda \phi \,,$ & 
\end{tabular}
\nonumber \\
&&
\begin{tabular}{l}
$\delta _\mu ^{-}\chi _{\nu \rho }=-D_\mu B_{\nu \rho }-i\theta _{\mu \nu
\rho \lambda }H^\lambda +i\theta _{\mu \nu \rho \lambda }D^\lambda \tau \,,$
\\ 
$\delta _\mu ^{-}\chi _\nu =4iH_{\mu \nu }+F_{\mu \nu }+4ig_{\mu \nu }\left[ 
\overline{\phi },\phi \right] -4\left[ B_{\mu \nu },\tau \right] \,,$ \\ 
$\delta _\mu ^{-}H_{\nu \rho }=D_\mu \psi _{\nu \rho }+\theta _{\mu \nu \rho
\lambda }\left( \left[ \psi ^\lambda ,\tau \right] -\left[ \chi ^\lambda
,\phi \right] -iD^\lambda \xi \right) +i\left[ \psi _\mu ,B_{\nu \rho
}\right] \,,$ \\ 
$\delta _\mu ^{-}H_\nu =-D_\mu \psi _\nu +D_\nu \psi _\mu +4\left[ \psi
_{\mu \nu },\tau \right] -4\left[ \xi ,B_{\mu \nu }\right] +4\left[ \chi
_{\mu \nu },\phi \right] +4ig_{\mu \nu }\left[ \eta ,\phi \right] \,.$%
\end{tabular}
\nonumber \\
&&  \label{VeTransf-}
\end{eqnarray}
where $\theta _{\mu \nu \rho \sigma }$ denotes the combination
\begin{equation}
\theta _{\mu \nu \rho \sigma }=\varepsilon _{\mu \nu \rho \sigma }+g_{\mu
\nu }g_{\rho \sigma }-g_{\mu \rho }g_{\nu \sigma }\;=4\Pi _{\mu \sigma \nu
\rho }^{+}\;,  \label{self}
\end{equation}
where $\Pi _{\mu \sigma \nu \rho }^{+}$ is the projector on self-dual
two-forms.

It is worth emphasizing that the invariant action $S_{N=4}$ is uniquely
fixed by the two vector generators $\delta _\mu ^{+},\;\delta _\mu ^{-}$ and
by the scalar charge $\delta ^{+}.$ In other words, the requirement of
invariance under $\delta _\mu ^{+},\;\delta _\mu ^{-}$ and $\delta ^{+%
\mathrm{\ }}$fixes all the relative numerical coefficients of the various
terms of the action $\left( \ref{invaction}\right) $. Due to this property,
the tensorial transformations $\delta _{\mu \nu }^{+}$, $\delta _{\mu \nu
}^{-}$ will not be taken into further account, although their inclusion can
be done straightforwardly. Notice that a similar situation has already been
met in the case of $N=2$ \cite{Sorella}.

Let us give here the algebraic relations among the twisted generators, 
\textit{i.e. }
\begin{eqnarray}
&&
\begin{tabular}{ll}
$\left\{ \delta ^{+},\delta ^{+}\right\} =\delta _{-2\phi }^g$ & $\left\{
\delta _\mu ^{+},\delta ^{+}\right\} =\partial _\mu +\delta _{A_\mu }^g$ \\ 
$\left\{ \delta ^{-},\delta ^{-}\right\} =\delta _{2\overline{\phi }}^g$ & $%
\left\{ \delta _\mu ^{-},\delta ^{-}\right\} =\partial _\mu +\delta _{A_\mu
}^g$ \\ 
$\left\{ \delta ^{+},\delta ^{-}\right\} =\delta _{-\tau }^g$ & $\left\{
\delta _\mu ^{+},\delta ^{-}\right\} =0$ \\ 
$\left\{ \delta _\mu ^{-},\delta ^{+}\right\} =0$ & $\left\{ \delta _\mu
^{+},\delta _\nu ^{+}\right\} =\delta _{-8g_{\mu \nu }\overline{\phi }}^g$
\\ 
$\left\{ \delta _\mu ^{-},\delta _\nu ^{-}\right\} =\delta _{8g_{\mu \nu
}\phi }^g$ & $\left\{ \delta _\mu ^{+},\delta _\nu ^{-}\right\} =\delta
_{-4iB_{\mu \nu }-4g_{\mu \nu }\tau }^g+\mathrm{eqs.\,\,of\,\,motion,}$%
\end{tabular}
\nonumber \\
&&  \label{DeltaAlgebra}
\end{eqnarray}
where $\delta _\gamma ^g$ denotes a gauge transformation with parameter $%
\gamma $.

\section{Quantization of the Twisted Theory}

In order to quantize the action $\left( \ref{invaction}\right) $ one has to
properly take into account the twisted generators $\delta ^{\pm },\delta
_\mu ^{\pm }.$ This amounts to perform the gauge fixing procedure in a way
compatible with the relevant global invariances of the action $S_{N=4}.$ Two
nontrivial aspects have therefore to be faced, namely: the nonlinearity of
the twisted transformations of the fields $\left( \ref{ScTransf+}\right)
-\left( \ref{VeTransf-}\right) $ and the fact that the algebra of the
generators $\delta ^{\pm },\delta _\mu ^{\pm }$ closes on the space-time
translations only on-shell and modulo field-dependent gauge transformations.
The way out in quantizing this kind of model requires the use of the
Batalin-Vilkovisky procedure, already successfully applied to the untwisted $%
N=4$ theory by \cite{White1}, and to $N=2$ by \cite{Maggiore,Sorella}.

The first step is to introduce the BRST symmetry corresponding to the gauge
invariance of the theory, \textit{i.e.}, $\delta _\zeta ^g\rightarrow s$ and 
$\zeta \rightarrow c$, where $c$ is the usual Faddeev-Popov (FP)\ ghost
transforming as $sc=ic^2$. Following now \cite{White1,Maggiore,Sorella}, one
extends the BRST operator into a new operator $Q$ which turns out to be
nilpotent on shell and which, together with the BRST charge $s$, collects
all the other generators appearing in the algebra $\left( \ref{DeltaAlgebra}%
\right) .$ Introducing thus a set of global ghosts $\omega ^{\pm
},\varepsilon ^{\pm \mu },v^\mu $ associated to $\delta ^{\pm },\delta _\mu
^{\pm },$ and to the space-time translations $\partial _\mu $, the operator $%
Q$ is found to be
\begin{equation}
Q=s+\omega ^{+}\delta ^{+}+\omega ^{-}\delta ^{-}+\varepsilon ^{+\mu }\delta
_\mu ^{+}+\varepsilon ^{-\mu }\delta _\mu ^{-}+v^\mu \partial _\mu -(\omega
^{+}\varepsilon ^{+\mu }+\omega ^{-}\varepsilon ^{-\mu })\frac \partial
{\partial v^\mu }\,.  \label{QOperator}
\end{equation}
>From the requirement that $Q$ carries FP charge $+1,$ it follows that the
global ghosts $\omega ^{+},\,\omega ^{-},\,\varepsilon ^{+\mu }$,$%
\,\varepsilon ^{-\mu }$ are commuting and have FP charge $+1$, while $v^\mu $
is anticommuting, with the same charge. Defining then the action of the
operator $Q$ on the Faddeev-Popov ghost $c$ as \cite{White1,Maggiore,Sorella}
\begin{eqnarray}
Qc &=&ic^2+(\omega ^{+^2}-4\varepsilon ^{-^2})\phi +(4\varepsilon
^{+^2}-\omega ^{-^2})\overline{\phi }+(\omega ^{+}\omega ^{-}+4\varepsilon
^{+\mu }\varepsilon _\mu ^{-})\tau  \nonumber \\
&&+4i\varepsilon ^{+\mu }\varepsilon ^{-\nu }B_{\mu \nu }-(\omega
^{+}\varepsilon ^{+\mu }+\omega ^{-}\varepsilon ^{-\mu })A_\mu \;,
\label{q-c}
\end{eqnarray}
the following equations are seen to hold
\begin{eqnarray}
QS_{N=4} &=&0\;,  \nonumber  \label{Qproperties} \\
Q^2 &=&0\,\,\,\,(\mathrm{modulo\,\,eqs.\,\,of\,\,motion})\,.
\label{Qproperties}
\end{eqnarray}
The next step is now to define the nonlinear transformations of the fields $%
Q\Phi _i$ by coupling them to antifields $\Phi _i^{*}$. This is done by
introducing the external action
\begin{equation}
S_{ext}=tr\int d^4x\;\Phi _i^{*}Q\Phi _i\;,  \label{sfontes}
\end{equation}
where, for a $p$-tensor field  

\[
\Phi _i^{*}Q\Phi _i=\frac 1{p!}\Phi _i^{*\mu _1..\mu _p}Q\Phi _{i\mu _1..\mu
_p}\;.
\]
In addition, due to the fact that the operator $Q$ is nilpotent only
on-shell, one has to introduce a further quadratic term in the antifields 
\cite{White1,Maggiore,Sorella}: 
\begin{eqnarray}
S_{quad} &=&4g^2\varepsilon ^{+\mu }\varepsilon ^{-\nu }tr\int d^4x\left(
\varepsilon _{\mu \nu \rho \lambda }A^{*\rho }H^{*\lambda }-\frac 12\left(
B_\nu ^{*\delta }H_{\mu \delta }^{*}-B_\mu ^{*\delta }H_{\nu \delta
}^{*}\right) \right.   \nonumber \\
&&\left. -\varepsilon _{\mu \nu \rho \lambda }\psi ^{*\rho }\chi ^{*\lambda
}+\frac 12\left( \psi _\nu ^{*\delta }\chi _{\mu \delta }^{*}-\psi _\mu
^{*\delta }\chi _{\nu \delta }^{*}\right) \right) \,.  \label{squad}
\end{eqnarray}
According to the Batalin-Vilkovisky procedure, we choose a Landau kind of
gauge fixing term which takes the following form
\begin{eqnarray}
S_{gf}=Q\,\,tr\int d^4x\left( \overline{c}\partial ^\mu A_\mu \right)
\,+\,4g^2\varepsilon ^{+\mu }\varepsilon ^{-\nu }\,tr\,\int d^4x\varepsilon
_{\mu \nu \rho \lambda }\partial ^\rho \overline{c}H^{*\lambda }\,,
\label{GFAction}
\end{eqnarray}
where the antighost $\overline{c},$ introduced by shifting the antifield $%
A_\mu ^{*}$ as $A_\mu ^{*}\rightarrow A_\mu ^{*}+\partial _\mu \overline{c}$%
, is required to transform as
\begin{eqnarray}
Q\overline{c} &=&b+v^\mu \partial _\mu \overline{c}\,,  \nonumber
\label{QAntTransf} \\
Qb &=&(\omega ^{+}\varepsilon ^{+\mu }+\omega ^{-}\varepsilon ^{-\mu
})\partial _\mu \overline{c}+v^\mu \partial _\mu b\,,  \label{QAntTransf}
\end{eqnarray}
$b$ being the Lagrange multiplier implementing the Landau condition. Let us
also display the quantum numbers of the ghosts and antifields
\begin{eqnarray*}
&& 
\begin{tabular}{|l|l|l|l|l|l|l|}
\hline
Field & $c$ & $\omega ^{+}$ & $\omega ^{-}$ & $\varepsilon ^{+\mu }$ & $%
\varepsilon ^{-\mu }$ & $v^\mu $ \\ \hline
G-ch. & $\,$0 & -1 & $\,\,\,$\thinspace 1 & $\,\,\,\,$1 & -1 & 0 \\ \hline
FP-ch. & $\,$1 & \thinspace \thinspace 1 & $\,\,\,\,$1 & $\,\,\,\,$1 & $\,\,$%
1 & $\,$1 \\ \hline
Dim. & \thinspace 0 & -$\frac 12$ & \thinspace \thinspace -$\frac 12$ & 
\thinspace \thinspace -$\frac 12$ & -$\frac 12$ & -1 \\ \hline
Nature & \thinspace a & \thinspace \thinspace c & \thinspace \thinspace c & 
\thinspace \thinspace \thinspace c & \thinspace \thinspace \thinspace c & 
\thinspace \thinspace a \\ \hline
\end{tabular}
\\
&&_{\,\,\,\,\,\,\,\,\,\,\,\mathrm{Table\,2.\,Quantum\,\,numbers\,\,of\,\,the\,%
\,ghost\,\,fields.}}\mathrm{\ \,}
\end{eqnarray*}
\begin{eqnarray*}
&& 
\begin{tabular}{|l|l|l|l|l|l|l|l|l|l|l|l|l|l|l|}
\hline
Source & $A_\mu ^{*}$ & $\xi ^{*}$ & $\eta ^{*}$ & $\psi _{\mu \nu }^{*}$ & $%
\chi _{\mu \nu }^{*}$ & $\psi _\mu ^{*}$ & $\chi _\mu ^{*}$ & $\phi ^{*}$ & $%
\tau ^{*}$ & $\overline{\phi }^{*}$ & $B_{\mu \nu }^{*}$ & $H_\mu ^{*}$ & $%
H_{\mu \nu }^{*}$ & $c^{*}$ \\ \hline
G-ch. & $\,\,$0 & -1 & $\,\,$1 & -1 & $\,$\thinspace 1 & -1 & $\,$\thinspace
1 & -2 & \thinspace \thinspace 0 & \thinspace \thinspace 2 & \thinspace
\thinspace 0 & \thinspace \thinspace 0 & \thinspace \thinspace 0 & 
\thinspace \thinspace 0 \\ \hline
FP-ch. & -1 & -1 & -1 & -1 & -1 & -1 & -1 & -1 & -1 & -1 & -1 & -1 & -1 & -2
\\ \hline
Dim. & $\,\,$3\thinspace \thinspace & $\,\,\frac 52$ & $\,\,\frac 52$ & $%
\,\,\frac 52$ & $\,\,\frac 52$ & $\,\,\frac 52$ & $\,\,\frac 52$ & $\,\,$3 & 
$\,\,$3 & $\,\,$3 & $\,\,$3 & $\,$\thinspace 2 & $\,\,$2 & $\,\,$4 \\ \hline
nature & \thinspace \thinspace a & \thinspace \thinspace c & \thinspace
\thinspace c & \thinspace \thinspace c & \thinspace \thinspace c & 
\thinspace \thinspace c & \thinspace \thinspace c & \thinspace \thinspace a
& \thinspace \thinspace a & \thinspace \thinspace a & \thinspace \thinspace a
& \thinspace \thinspace a & \thinspace \thinspace a & \thinspace \thinspace c
\\ \hline
\end{tabular}
\\
&&_{\,\,\,\,\,\,\,\,\,\,\,\,\,\,\,\,\,\,\,\,\,\,\,\,\,\,\,\,\,\,\,\,\,\,\,\,%
\,\,\,\,\,\,\,\,\,\,\,\,\,\,\,\,\,\,\,\,\,\,\,\,\,\,\,\,\,\,\,\,\,\,\,\,\,\,%
\,\,\,\,\,\,\,\,\,\,\,\mathrm{Table\,3.\,Quantum\,\,numbers\,\,of\,\,the\,%
\,antifields.}}
\end{eqnarray*}
Finally, the complete gauge-fixed action $\Sigma $
\begin{equation}
\Sigma =S_{N=4}+S_{ext}+S_{quad}+S_{gf}\,\,,  \label{TotAction}
\end{equation}
turns out to obey the following Slavnov-Taylor identity
\begin{equation}
\mathcal{S}\left( \Sigma \right) =0\;,  \label{st0}
\end{equation}
with
\begin{eqnarray}
\mathcal{S}\left( \Sigma \right)  &=&tr\int d^4x\left( \frac{\delta \Sigma }{%
\delta \Phi _i^{*}}\frac{\delta \Sigma }{\delta \Phi _i}+\left( b+v^\mu
\partial _\mu \overline{c}\right) \frac{\delta \Sigma }{\delta \overline{c}}%
-(\omega ^{+}\varepsilon ^{+\mu }+\omega ^{-}\varepsilon ^{-\mu })\frac{%
\delta \Sigma }{\delta v^\mu }\right.   \nonumber \\
&&\left. +\left( (\omega ^{+}\varepsilon ^{+\mu }+\omega ^{-}\varepsilon
^{-\mu })\partial _\mu \overline{c}+v^\mu \partial _\mu b\,\right) \frac{%
\delta \Sigma }{\delta b}\right) \;.  \label{stop0}
\end{eqnarray}
We underline that the Slavnov-Taylor identity $\left( \ref{st0}\right) $
contains all the information concerning both the gauge and the $N=4$
invariances of the classical starting action $S_{N=4}$. In other words, the
present algebraic set up has allowed to obtain a gauge-fixed action which is
compatible with the full set of defining symmetries. This point is of
particular relevance, as allows one to analyse the quantum aspects of the
theory by means of the powerful BRST\ cohomology techniques \cite{book}.
This will be the goal of the next section. However, before going any
further, it is convenient to eliminate the ghost of the space-time
translations $v_\mu $ by introducing the so-called reduced action $%
\widetilde{S}$
\begin{equation}
\Sigma \rightarrow \widetilde{S}=\Sigma -v^\rho \frac{\partial \Sigma }{%
\partial v^\rho }-tr\int d^4x\,b\partial ^\mu A_\mu \;.  \label{r-ac}
\end{equation}
In terms of the action $\widetilde{S}$, the Slavnov-Taylor identity takes
now the simplified form
\begin{equation}
\mathcal{S}\left( \widetilde{S}\right) =tr\int d^4x\left( \frac{\delta 
\widetilde{S}}{\delta \Phi _i^{*}}\frac{\delta \widetilde{S}}{\delta \Phi _i}%
\right) =(\omega ^{+}\varepsilon ^{+\mu }+\omega ^{-}\varepsilon ^{-\mu
})\Delta _\mu ^{cl}  \label{SlavReduc1}
\end{equation}
where the antifield $A^{*\mu }$ has to be replaced by the shifted antifield $%
\widehat{A}_\mu ^{*}=A_\mu ^{*}+\partial _\mu \overline{c}$, and $\Delta
_\mu ^{cl}$ is given by
\begin{eqnarray}
\Delta _\rho ^{cl} &=&tr\int d^4x\left( -\widehat{A}_\mu ^{*}\partial _\rho
A^\mu -H^{*\mu }\partial _\rho H_\mu -\frac 12B^{*\mu \nu }\partial _\rho
B_{\mu \nu }-\tau ^{*}\partial _\rho \tau \right.   \nonumber
\label{sfontes} \\
&&-\frac 12H^{*\mu \nu }\partial _\rho H_{\mu \nu }++\frac 12\psi ^{*\mu \nu
}\partial _\rho \psi _{\mu \nu }+\frac 12\chi ^{*\mu \nu }\partial _\rho
\chi _{\mu \nu }+\psi ^{*\mu }\partial _\rho \psi _\mu \;  \nonumber \\
&&\left. \frac {}{}+\chi ^{*\mu }\partial _\rho \chi _\mu +\xi ^{*}\partial
_\rho \xi +\eta ^{*}\partial _\rho \eta -\phi ^{*}\partial _\rho \phi -%
\overline{\phi }^{*}\partial _\rho \overline{\phi }+c^{*}\partial _\rho
c\right) \,.  \label{que}
\end{eqnarray}
Notice that the above expression, being linear in the quantum fields, is a
classical breaking and will not be affected by the quantum corrections. In
addition, the Slavnov-Taylor identity $\left( \ref{SlavReduc1}\right) $
implies that the linearized Slavnov-Taylor operator $B_{\widetilde{S}}$ \cite
{book} defined as
\begin{equation}
B_{\widetilde{S}}=tr\int d^4x\left( \frac{\delta \widetilde{S}}{\delta \Phi
_i^{*}}\frac \delta {\delta \Phi _i}+\frac{\delta \widetilde{S}}{\delta \Phi
_i}\frac \delta {\delta \Phi _i^{*}}\right) \;,  \label{LinSlavReduc1}
\end{equation}
has the following property 
\begin{equation}
B_{\widetilde{S}}B_{\widetilde{S}}=(\omega ^{+}\varepsilon ^{+\mu }+\omega
^{-}\varepsilon ^{-\mu })\partial _\mu \,\;,  \label{LinSlavReduc2}
\end{equation}
which shows that $B_{\widetilde{S}}$ is nilpotent modulo a total derivative.
As we shall see in detail in the next section, this feature will give rise
to a set of very constrained descent equations which will allow to relate
the full action of $N=4$ to the gauge invariant polynomial $tr\,\phi ^2$, $%
\phi $ being the scalar field of the $N=2$ vector multiplet.

\section{The relationship between the action of N=4 and $tr\,\phi ^2$}

In order to establish the relationship between the action of $N=4$ and $%
tr\,\phi ^2$ we proceed by analysing the cohomology of the operator $B_{%
\widetilde{S}}$ in the space of the integrated local functionals of the
fields and their derivatives, with the same quantum numbers of the classical
action. To this end it is worth recalling that the invariant action $S_{N=4}$
is uniquely fixed by the three generators $\delta _\mu ^{+},\;\delta _\mu
^{-}$ and $\delta ^{+}.$ We can thus simplify our analysis setting to zero
the global parameter $\omega ^{-}$ corresponding to the scalar charge $%
\delta ^{-}.$ Therefore, eqs.$\left( \ref{SlavReduc1}\right) $ and $\left( 
\ref{LinSlavReduc2}\right) $ take the following form
\begin{eqnarray}
\mathcal{S}\left( \widetilde{S}\right) &=&\omega ^{+}\varepsilon ^{+\mu
}\Delta _\mu ^{cl}\,,  \label{SlavReducMod} \\
B_{\widetilde{S}}B_{\widetilde{S}} &=&\omega ^{+}\varepsilon ^{+\mu
}\partial _\mu \,.  \label{LinSlavReducMod}
\end{eqnarray}
It should be observed from $\left( \ref{LinSlavReducMod}\right) $ that the
operator $B_{\widetilde{S}}$ is not strictly nilpotent, as its square yields
the space-time translations. This property is indeed a general feature of
the supersymmetric gauge theories, following from the fact that the
supersymmetric algebra closes on the space-time translations. In particular,
equation $\left( \ref{LinSlavReducMod}\right) $ will give rise to a set of
descent equations which will strongly constrain the cohomology of $B_{%
\widetilde{S}}$. As one can expect, this is related to the large number of
generators which are encoded in the operator $B_{\widetilde{S}}$ ,
reflecting the $N=4$ structure of the theory.

In order to obtain the descent equations for the operator $B_{\widetilde{S}} 
$ , let us start with the consistency condition
\begin{equation}
B_{\widetilde{S}}\int d^4x\,\Omega ^0=0\;,  \label{w-z}
\end{equation}
where $\Omega ^0$ has the same quantum numbers of the classical Lagrangian
of $N=4$, \textit{i.e.} it is a local polynomial of dimension four and with
vanishing FP and G-charge. Due to eq.$\left( \ref{LinSlavReducMod}\right) $,
the integrated consistency condition $\left( \ref{w-z}\right) $ can be
translated at the local level as
\begin{equation}
B_{\widetilde{S}}\Omega ^0=\partial ^\mu \Omega _{\mu \;}^1,  \label{1-l}
\end{equation}
where $\Omega _{\mu \;}^1$is a local polynomial of FP charge 1 and dimension
3. Applying now the operator $B_{\widetilde{S}}$ to both sides of $\left( 
\ref{1-l}\right) $ and making use of eq.$\left( \ref{LinSlavReducMod}\right) 
$, one obtains the condition
\begin{equation}
\partial ^\mu \left( B_{\widetilde{S}}\Omega _{\mu \;}^1-\omega
^{+}\varepsilon _\mu ^{+}\Omega ^0\right) =0\;,  \label{l-2}
\end{equation}
which, due to the algebraic Poincar\'{e} Lemma \cite{book}, implies
\begin{equation}
B_{\widetilde{S}}\Omega _{\mu \;}^1=\omega ^{+}\varepsilon _\mu ^{+}\Omega
^0+\partial ^\nu \Omega _{[\nu \mu ]}^2\;,  \label{l-3}
\end{equation}
for some local polynomial $\Omega _{[\nu \mu ]}^2$ antisymmetric in the
Lorentz indices $\mu ,\nu $ and with FP charge 2. The procedure can be
easily iterated, yielding the following set of descent equations
\begin{eqnarray}
B_{\widetilde{S}}\Omega ^0 &=&\partial ^\mu \Omega _\mu ^1\;,  \nonumber \\
\,\,\,\,\,\,\,\,\;\;\;\;\;B_{\widetilde{S}}\Omega _\mu ^1 &=&\partial ^\nu
\Omega _{[\nu \mu ]}^2+\omega ^{+}\varepsilon _\mu ^{+}\Omega ^0\;, 
\nonumber \\
B_{\widetilde{S}}\Omega _{[\mu \nu ]}^2 &=&\partial ^\rho \Omega _{[\rho \mu
\nu ]}^3+\omega ^{+}\varepsilon _\mu ^{+}\Omega _\nu ^1-\omega
^{+}\varepsilon _\nu ^{+}\Omega _\mu ^1\;,  \nonumber \\
B_{\widetilde{S}}\Omega _{[\mu \nu \rho ]}^3 &=&\partial ^\sigma \Omega
_{[\sigma \mu \nu \rho ]}^4+\omega ^{+}\varepsilon _\mu ^{+}\Omega _{[\nu
\rho ]}^2+\omega ^{+}\varepsilon _\rho ^{+}\Omega _{[\mu \nu ]}^2+\omega
^{+}\varepsilon _\nu ^{+}\Omega _{[\rho \mu ]}^2,  \nonumber \\
B_{\widetilde{S}}\Omega _{[\mu \nu \rho \sigma ]}^4 &=&\omega
^{+}\varepsilon _\mu ^{+}\Omega _{[\nu \rho \sigma ]}^3-\omega
^{+}\varepsilon _\sigma ^{+}\Omega _{[\mu \nu \rho ]}^3-\omega
^{+}\varepsilon _\rho ^{+}\Omega _{[\sigma \mu \nu ]}+\Omega _{[\rho \sigma
\mu ]}^3\omega ^{+}\varepsilon _\nu ^{+}\,.  \nonumber  \label{DesEqB} \\
&&  \label{DesEqB}
\end{eqnarray}
It is interesting to observe that these equations are of an unusual type, as
the cocycles with low FP charge appear in the equations of those with higher
FP charge, turning the system $\left( \ref{DesEqB}\right) $ highly
nontrivial. We remark that the last equation for $\Omega _{[\mu \nu \rho
\sigma ]}^4$ is not homogeneous, a property which strongly constrains the
possible solutions. To some extent, the equations $\left( \ref{DesEqB}%
\right) $ display a certain similarity with the descent equations in $N=1$
superspace \cite{Piguet}. Actually, it is possible to solve the system $%
\left( \ref{DesEqB}\right) $ in a rather direct way by making use of the $N=4
$ structure. To this end, let us introduce the operator
\begin{equation}
\mathcal{W}_\mu =\frac 1{\omega ^{+}}\left[ \frac \partial {\partial
\varepsilon ^{+\mu }},B_{\widetilde{S}}\right] \;,  \label{ClimbDesEq}
\end{equation}
which obeys the relations
\begin{eqnarray}
\left\{ \mathcal{W}_\mu ,B_{\widetilde{S}}\right\}  &=&\partial _\mu \,, 
\nonumber  \label{EBAlgebra} \\
\left\{ \mathcal{W}_\mu ,\mathcal{W}_\nu \right\}  &=&0\,\;\mathrm{.}
\label{EBAlgebra}
\end{eqnarray}
This algebra is typical of  topological quantum field theories \cite{Delduc}%
. In particular, as shown in \cite{Guadagnini}, the decomposition $\left( 
\ref{EBAlgebra}\right) $ allows to making use of $\mathcal{W}_\mu $ as a
climbing-up operator for the descent equations $\left( \ref{DesEqB}\right) .$
It turns out in fact that the solution of the system is 
\begin{eqnarray}
\Omega ^0 &=&\frac 1{4!}\mathcal{W}^\mu \mathcal{W}^\nu \mathcal{W}^\rho 
\mathcal{W}^\sigma \Omega _{[\sigma \rho \nu \mu ]}^4\;,  \nonumber
\label{sol} \\
\Omega _\mu ^1 &=&\frac 1{3!}\mathcal{W}^\nu \mathcal{W}^\rho \mathcal{W}%
^\sigma \Omega _{[\sigma \rho \nu \mu ]}^4\;,  \nonumber \\
\Omega _{[\mu \nu ]}^2 &=&\frac 1{2!}\mathcal{W}^\rho \mathcal{W}^\sigma
\Omega _{[\sigma \rho \mu \nu ]}^4\;,  \nonumber \\
\Omega _{[\mu \nu \rho ]}^3 &=&\mathcal{W}^\sigma \Omega _{[\sigma \mu \nu
\rho ]}^4\;,  \label{sol}
\end{eqnarray}
with $\Omega _{[\mu \nu \rho \sigma ]}^4$ given by
\begin{equation}
\Omega _{[\mu \nu \rho \sigma ]}^4=\left( \omega ^{+}\right) ^4\varepsilon
_{\mu \nu \rho \sigma }tr\,\phi ^2.  \label{phi2}
\end{equation}
>From eqs.$\left( \ref{sol}\right) $ the usefulness of the operator $\mathcal{%
W}_\mu $ becomes now apparent. Recalling thus that the cocycle $\Omega ^0$
has the same quantum numbers of the $N=4$ Lagrangian, the following relation
holds
\begin{eqnarray}
\left( g\frac{\partial \widetilde{S}}{\partial g}-\varepsilon ^{-\mu }\frac{%
\partial \widetilde{S}}{\partial \varepsilon ^{-\mu }}\right) _{\omega
^{-}=0} &=&-\frac{\left( \omega ^{+}\right) ^4}{96g^2}\varepsilon ^{\mu \nu
\rho \sigma }\mathcal{W}_\mu \mathcal{W}_\nu \mathcal{W}_\rho \mathcal{W}%
_\sigma tr\int d^4x\,\phi ^2\;+\;B_{\widetilde{S}}\Xi ^{-1}\;,  \nonumber \\
&&  \label{RelActionLastLevel}
\end{eqnarray}
for some irrelevant trivial cocycle $\Xi ^{-1}.\;$Eq.$\left( \ref
{RelActionLastLevel}\right) $ follows from the observation that the
derivatives of the Slavnov-Taylor identity $\left( \ref{SlavReducMod}\right) 
$ with respect to the coupling constant $g$ and to the global ghost $%
\varepsilon ^{-\mu }$ define integrated cohomology classes of the operator $%
B_{\widetilde{S}}$, namely
\begin{equation}
B_{\widetilde{S}}\left( \frac{\partial \widetilde{S}}{\partial g}\right)
_{\omega ^{-}=0}=0\;,  \label{coup}
\end{equation}
and
\begin{equation}
B_{\widetilde{S}}\left( \frac{\partial \widetilde{S}}{\partial \varepsilon
^{-\mu }}\right) _{\omega ^{-}=0}=0\;,  \label{n-t}
\end{equation}
the first equation establishing the coupling constant $g$ as a physical
parameter of the theory. 

We remark that the presence of the second term in the lhs of eq.$\left( \ref
{RelActionLastLevel}\right) $ is a consequence of setting to zero the ghost
parameter $\omega ^{-}$, which implies that the classical breaking term in
the rhs of the Slavnov-Taylor identity $\left( \ref{SlavReduc1}\right) $
becomes independent from $\varepsilon ^{-\mu }.\;$Furthermore, this term has
a precise meaning, being related to the $\delta _{\mu \text{ }}^{-}-$%
invariance of the action $\left( \ref{invaction}\right) $. In fact, taking
the equation $\left( \ref{n-t}\right) $ at zero antifields and ghosts, one
gets
\begin{equation}
B_{\widetilde{S}}\left( \frac{\partial \widetilde{S}}{\partial \varepsilon
^{-\mu }}\right) _{\Phi ^{*},\;\mathrm{ghosts}=0}=\delta _\mu
^{-}S_{N=4}=0\;.  \label{c-n}
\end{equation}
In particular, setting also $\varepsilon ^{-\mu }=0$ in the eq.$\left( \ref
{RelActionLastLevel}\right) $, we finally obtain the key relationship
\begin{equation}
\left( g\frac{\partial \widetilde{S}}{\partial g}\right) _{\omega
^{-},\varepsilon ^{-}=0}\approx -\frac{\left( \omega ^{+}\right) ^4}{96g^2}%
\varepsilon ^{\mu \nu \rho \sigma }\mathcal{W}_\mu \mathcal{W}_\nu \mathcal{W%
}_\rho \mathcal{W}_\sigma tr\int d^4x\,\phi ^2\;  \label{fund}
\end{equation}
where $\approx $ means that the above equation holds modulo a trivial
cocycle. We see therefore that, as announced, the full gauge-fixed action of 
$N=4$ can be traced back to the gauge invariant polynomial $tr\,\phi ^2.$ It
should be noted that the introduction of the global ghosts $\omega ^{\pm
},\varepsilon _\mu ^{\pm }$ associated to the generators of $N=4$ has to be
seen as a useful device for carrying out a quantization procedure compatible
with all invariances of the starting action. Since the physical observables
of the theory, as for instance the chiral primary operators, are just
required to be gauge invariant, the global ghosts have to be set to zero at
the end.

The relation $\left( \ref{fund}\right) $ represents the main result of the
present work. It implies that many properties of the $N=4$ action can be
understood by looking at the gauge invariant polynomial $tr\,\phi ^2,$ which
plays the r\^{o}le of a kind of perturbative prepotential. In particular,
following the same algebraic procedure adopted in the case of $N=2$ \cite
{Blasi}, it is easy to prove that the anomalous dimension of the operator $%
tr\,\phi ^2$ vanishes to all orders of perturbation theory, implying that
the $\beta -$function of the $N=4$ can be at most of the order one-loop \cite
{Blasi}. However, a direct inspection of the one-loop coefficient shows that 
$\beta $ vanishes to all orders of perturbation theory. Consider in fact the
general expression for the one-loop $\beta -$function of $N=2$ \cite
{Weinberg}
\begin{equation}
\beta (g)=-\frac{g^2}{8\pi ^2}\left( C_1-hC_2\right) \;,
\label{betafunction}
\end{equation}
where $C_1$ and $C_2$ are respectively the Casimir invariants of the
representations of gauge and matter $N=2$ multiplets, $h$ being the number
of matter multiplets. As is well known, the $N=4$ SYM can be obtained by $N=2
$ when $h=1$ and the matter multiplet belongs to the adjoint representation.
Therefore $C_1=C_2$, implying that $\beta (g)=0$. This simple understanding
of the perturbative ultraviolet finiteness of the theory provides a
nontrivial example of how the properties of $N=4$ can be in fact understood
in terms of those of $tr\,\phi ^2$.

\section{Conclusion}

We have seen that the use of the twisted version of $N=4$ allows to identify
a subset of supercharges actually relevant for the nonrenormalization
aspects of the theory. This subset is given by the generators $\delta
^{+},\delta _\mu ^{+},\delta _\mu ^{-},$ which uniquely fix the classical
action. In addition, the vector charge $\delta _\mu ^{+}$ allows to solve
easily the descent equations, leading us to relate the gauge-fixed action of 
$N=4$ to the operator $tr\phi ^2$, see eq.$\left( \ref{fund}\right) $. This
result can be understood as an off-shell extension of the reduction formula
discussed in \cite{KI}, which plays an important r\^{o}le in the analysis of
the nonrenormalization theorems for the correlators of chiral primary
operators \cite{skiba,Sokatchev}.

Moreover, as proven in \cite{Blasi}, the Ward identity associated to the
scalar supersymmetry  implies the vanishing of the anomalous dimension of $%
tr\,\phi ^2$. This result may be considered as a first step towards an
all-orders BRST algebraic analysis of the scaling properties of the above
mentioned correlators.

It would be also useful to generalize the eq.$\left( \ref{fund}\right) $ to
the case of a nonvanishing central charge for the $N=4$ superalgebra, which would
allow us to extend the proof of the ultraviolet finiteness of the theory to
the spontaneously broken phase \cite{pr}.

\section*{Acknowledgements}

The Conselho Nacional de Desenvolvimento Cient\'{\i}fico e Tecnol\'{o}gico
CNPq-Brazil, the Funda\c {c}\~{a}o de Amparo \`{a} Pesquisa do Estado do Rio
de Janeiro (Faperj) and the SR2-UERJ are acknowledged for the financial
support. A.T. thanks the Departamento de F\'{\i}sica Te\'{o}rica of the
UERJ\ for the kind hospitality.

\end{document}